\documentclass[journal,twocolumn]{IEEEtran}
\usepackage{epsfig,makeidx,color}
\usepackage{amsmath,amssymb,amsthm,bbm}
\usepackage{cite,graphicx}
\usepackage{enumerate}
\usepackage{hyperref}
\hypersetup{
        colorlinks = true,
        citecolor=blue,
}

\def\cA{{\cal A}}

\def\rH{{\rm H}}

\def\rT{{\rm T}}

\def\uE{{\mathbb E}}

\DeclareMathOperator*{\argmin}{\arg\!\min}

\def\indicator{\mathbbm{1}}

\def\be{ \begin{equation} }
\def\ee{ \end{equation} }
\def\bea{ \begin{eqnarray} }
\def\eea{ \end{eqnarray} }
\def\bx{{\bf x}}

\def\bs{{\bf s}}
\def\ba{{\bf a}}
\def\br{{\bf r}}
\def\bu{{\bf u}}

\def\bn{{\bf n}}

\def\bz{{\bf z}}

\def\bA{{\bf A}}

\def\bI{{\bf I}}

\def\bR{{\bf R}}

\def\b0{{\bf 0}}

\def\bPsi{{\bf \Psi}}

\def\bpsi{{\boldsymbol \psi}}

\def\cB{{\cal B}}
\def\cC{{\cal C}}
\def\cD{{\cal D}}

\def\cN{{\cal N}}

\ifCLASSOPTIONonecolumn
  \newcommand{\figwidth}{0.60\columnwidth}
\else
  \newcommand{\figwidth}{0.90\columnwidth}
\fi

\begin{document}

\title{Gaussian Data-aided Sensing with Multichannel Random
Access and Model Selection}

\author{Jinho Choi
\thanks{The author is with
the School of Information Technology,
Deakin University, Geelong, VIC 3220, Australia
(e-mail: jinho.choi@deakin.edu.au)}}

\maketitle

\begin{abstract}
In this paper, we study data-aided sensing 
(DAS) for a system consisting of a base station (BS) and a number of
nodes, where the BS becomes a receiver that collects
measurements or data sets from the nodes that are distributed over a cell.
DAS is an iterative data collection scheme 
that allows the BS to
efficiently estimate a target signal 
(i.e., all nodes' measurements)
with a small number of measurements (compared to random polling).
In DAS, a set of nodes are selected in each round
based on the data sets that are already available at the BS
from previous rounds for efficient data collection.
We consider DAS for measurements
that are correlated Gaussian in this paper. The resulting DAS
is referred to as Gaussian DAS.
Using the mean squared error (MSE) criterion,
in each round, the BS is able to choose a node
that has a data set to minimize the MSE of the next round.
Furthermore, we generalize Gaussian DAS in two different ways: i) 
with multiple parallel channels to upload measurements from nodes 
using random access; ii) with a model selection,
where a multi-armed bandit problem formulation
is
used to combine the model selection with DAS.
\end{abstract}

{\IEEEkeywords
Internet of Things; 
Intelligent Data Collection;
Model Selection; Multi-armed Bandit}

\ifCLASSOPTIONonecolumn
\baselineskip 26pt
\fi

\section{Introduction} \label{S:Intro}

The Internet of Things (IoT) has been 
an important issue as it has a number of applications in
various areas including smart cities and factories
in the future
\cite{Gubbi13} 
\cite{Kim16}.
To build IoT systems, 
layered approaches are usually considered,
where the bottom layer is responsible for collecting and processing
information or data from devices or sensors
\cite{Fuqaha15}.

Cellular IoT has been considered to support IoT applications
over a large area. For example, in \cite{Mang16},
a deployment study of narrowband IoT (NB-IoT) 
\cite{3GPP_NBIoT} is carried out
to support IoT applications over a large area.
In cellular IoT (for the bottom layer
in IoT systems), each base station (BS)
can be used as a data collector from devices
or sensors deployed over a cell.
Since long-term evolution (LTE)
BSs are well deployed, cellular IoT 
might play a crucial role as IoT infrastructure in collecting
a large amount of data from devices including mobile phones over
a wide area.

Collecting data sets from devices deployed in an area
requires devices' sensing to acquire local measurements or data 
and uploading to a BS in cellular IoT.
While sensing and uploading can be considered separately,
they can also be combined, which leads to
data-aided sensing (DAS) \cite{Choi19}.
DAS is an iterative data collection scheme
where a BS is to collect data sets from devices or nodes
(we use devices and nodes interchangeably)
through multiple rounds.
In DAS, the BS chooses a set of nodes in each round
based on the data sets that are already available at the BS
from previous rounds for efficient data collection.
As a result, 
the BS is able to 
efficiently estimate a target signal 
(i.e., all nodes' measurements)
with a small number of measurements compared to random polling.

In this paper, 
we consider DAS when measurements
at nodes are modeled as correlated Gaussian random variables. 
Note that in 
\cite{Choi19}, it is assumed that measurements
have a sparse representation 
so that the notion of compressive sensing (CS)
\cite{Donoho06} \cite{Candes06} is exploited.
While measurements at nodes can have a sparse
representation as in \cite{Choi19}, 
sensor nodes in wireless sensor networks
may observe correlated Gaussian sources 
as in
\cite{Vuran04} \cite{Bachceci08} \cite{Chen12}. 
Thus, it would be necessary to apply DAS to correlated 
Gaussian signals. 
Since the measurements are Gaussian,
they can be characterized by the mean
vector and covariance matrix.
Thus, in this paper, we assume
that the mean vector and covariance matrix
of nodes' measurements 
are available at the BS. From them,
based on the minimum mean squared error (MMSE) criterion,
the BS can perform DAS to effectively
collect data sets or measurements from nodes.
Thanks to the Gaussian assumption for measurements,
a closed-form expression that allows 
the BS to decide 
the best node in each round 
is available based on the MMSE criterion.

We also generalize DAS
in this paper in two different ways.
First, multiple parallel channels
are considered to collect data sets from nodes.
In this case, the BS can have more data
sets from multiple selected nodes in each round. 
However, some nodes may experience
deep fading or do not have measurements
yet for various reasons. In this case,
some of multiple parallel channels are not 
utilized when the associated 
nodes cannot transmit their measurements,
which leads to a low utilization of multiple parallel channels.
To mitigate this problem,
multichannel random access,
e.g., multichannel ALOHA \cite{Shen03} \cite{Chang15},
can be used.
In particular, when the probability
that a requested node can transmit its measurement
is sufficiently low (e.g., less than $e^{-1}$),
we can show that multichannel ALOHA can provide 
better performance than sequential polling with 
the number of nodes selected per round, 
which is greater than the number of parallel channels.

Secondly, we consider the case that there are multiple
models for measurements. In this case,
without knowing the correct model in advance,
the BS needs to perform DAS.
We show that the problem can be seen
as a multi-armed bandit problem \cite{Sutton98}
\cite{Bubeck12}.
Each model is seen as a one-armed bandit machine 
and the BS needs to explore all models
before it chooses a model as the correct one
(for exploitation) in DAS to collect data from
nodes.

The rest of the paper is organized as follows.
In Section~\ref{S:SM}, we present the system model
for DAS.
In Section~\ref{S:DAS}, for Gaussian measurements, 
DAS is studied based on the MMSE criterion.
Gaussian DAS is generalized with
multiple parallel channels in Section~\ref{S:MRA},
where random access is also considered for uploading.
A model selection problem is studied in conjunction with DAS
in Section~\ref{S:MAB}, which is seen as a multi-armed bandit problem.
We present simulation results in Section~\ref{S:Sim}
and conclude the paper with remarks
in Section~\ref{S:Conc}.

{\it Notation}:
Matrices and vectors are denoted by upper- and lower-case
boldface letters, respectively.
The superscript $\rT$
denotes the transpose and ${\rm Tr}(\bA)$
represents the trace of a square matrix $\bA$.
$\uE[\cdot]$
and ${\rm Var}(\cdot)$
denote the statistical expectation and variance, respectively.
In addition, ${\rm Cov}(\bx)$ represents
the covariance matrix of random vector $\bx$.
$\cN(\ba, \bR)$
and 
$\cC\cN(\ba, \bR)$
represent the distributions of
real-valued Gaussian 
and circularly symmetric complex Gaussian (CSCG)
random vectors with mean vector $\ba$ and
covariance matrix $\bR$, respectively.

\section{System Model} \label{S:SM}

\begin{figure}[thb]
\begin{center}
\includegraphics[width=\figwidth]{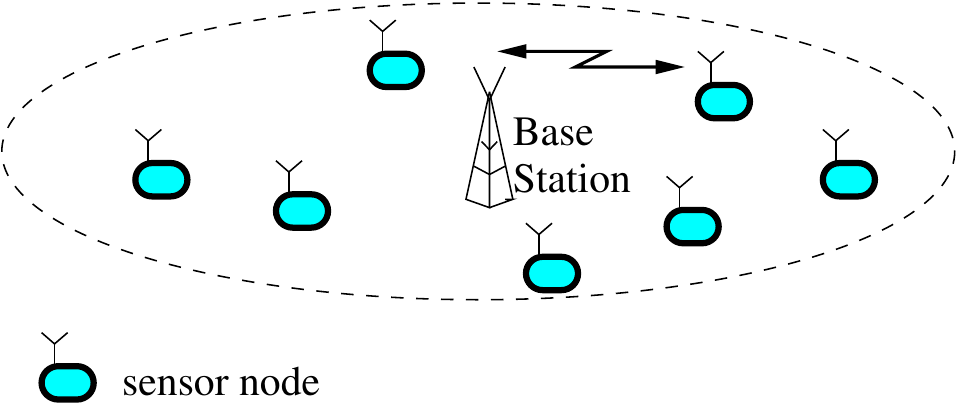}
\end{center}
\caption{Sensor nodes deployed in a certain area to collect
local data sets and upload to a base station.}
        \label{Fig:cell}
\end{figure}

Suppose that there are $K$ sensor nodes that
are deployed over a certain area so that each node can 
collect local environmental data. 
There is a receiver node, which is assumed to
an access point (AP) or BS to collect data sets
from nodes as illustrated in Fig.~\ref{Fig:cell}.
Thus, with an infrastructure-based wireless local
area network (WLAN), 
the system illustrated in Fig.~\ref{Fig:cell}
can be implemented.

Denote by $x_k$ the measurement 
at node $k$. In this
paper, we use measurements and data (sets) interchangeably.
Note that the measurement at each node can be a vector. However,
for simplicity, we assume that $x_k$ is a real-valued scalar
in this paper.
For convenience, let
$$
\bx = [x_1 \ \ldots \ x_K]^\rT,
$$
which is referred to as the target signal.
We assume that the BS is to obtain $\bx$ or its estimate.
To this end, all $K$ nodes are to transmit their data
sets to the BS.
For example, polling \cite{BertsekasBook} can be used where 
all $K$ nodes transmit their measurements to the BS 
sequentially in a specific order.
When the measurements 
$x_1, \ldots, x_K$ are independent of each other, 
all the $K$ nodes need to upload their measurements.
On the other hand, if they are correlated,
a good estimate of $\bx$ can be obtained
at the BS from the measurements uploaded by
a subset of $K$ nodes using the correlation
(i.e., a small subset of $\bx$ can be sufficient to
find a good estimate of $\bx$ if the $x_k$'s are highly correlated).

\section{DAS based on MMSE Criterion}	\label{S:DAS}

In order to exploit the correlation between
nodes' measurements in finding
an estimate of $\bx$ 
or reconstructing the target signal
with a small number of nodes transmitting
their measurements, 
in \cite{Choi19}, the notion of DAS is proposed when $\bx$
has a sparse representation.
In this section, we consider DAS in a different setting
based on the
MMSE
criterion.

Suppose that the BS is
to collect the measurements of nodes 
in multiple rounds by polling.
In addition, we assume that only one
node is able to transmit its measurement 
at each time, and any node is ready to transmit
its measurement up on request from the BS and
its transmission is always successful
in this section 
(these assumptions will be relaxed
later).

Let $t$ be the index for rounds.
At round $t = 0$, suppose that the BS chooses 
a node and sends the request signal
to upload its measurement.
For convenience,
suppose that $\bx$ is divided into two subvectors, $\bu_t$ and $\bz_t$,
where $\bz_t$ 
and $\bu_t$ are the subvectors of $\bx$ that are available
and unavailable at the BS, respectively,
at round $t$. 
Clearly, the lengths of $\bz_t$ 
and $\bu_t$ are $t$ and $K-t$, respectively.

For convenience, we omit the round index $t$.
Since the BS has $\bz$, the mean squared error (MSE)
of $\bx$ becomes 
$\uE[||\bu  - \hat \bu(\bz)||^2]$,
where $\hat \bu(\bz)$ is an estimate of $\bu$ with known 
$\bz$.
In the next round, an element of $\bu$ might be available
as the BS is to choose one of the nodes associated with $\bu$.
For convenience, let $L$ denote the length of $\bu$.
Among $L$ nodes, it is expected to choose the node
that can minimize the MSE. 
To this end, suppose that $\bu$ is further divided into
$u_l$ and $\bu_{-l}$,
where $u_l$ 
represents the $l$th element of $\bu$ and
$$
\bu_{-l} = [u_1 \ \ldots \ u_{l-1} \ u_{l+1} \ \ldots \ u_L]^\rT.
$$
Define the MSE  
when $u_l$ is to be available as
\be
C_l = \uE[||\bu_{-l}  - \hat \bu_{-l}(\bz, u_l)||^2],
	\label{EQ:C_l}
\ee
where $\hat \bu_{-l}(\bz, u_l)$ is the MMSE estimator
for $\bu_{-l}$,
which is the conditional mean of $\bu_{-l}$ \cite{Anderson79}
\cite{ChoiJBook2}, i.e.,
$$
\hat \bu_{-l}(\bz, u_l) =  \uE[\bu_{-l}\,|\, \bz, u_l].
$$
Then, in the next round,
the node to transmit its data to the BS provided that
the BS has $\bz$ can be chosen as follows:
\be
l^* = \argmin_{l \in \{1,\ldots,L\}} C_l
\ee
to minimize the MSE,
which in turn leads to a good estimate of $\bx$
with a smaller number of rounds
(than that using random polling).

Note that if $\bx$ is Gaussian,
the MMSE estimator is a linear estimator \cite{Anderson79}.
Thus, $C_l$ can also be expressed as
\be
C_l = \min_{\ba_l} \uE[||\tilde \bu_{-l} - 
\ba_l \tilde u_l||^2 \, \bigl|\, \bz],
\ee
where
$\tilde \bu_{-l} = \bu_{-l} - \uE[\bu_{-l}\,|\, \bz]$,
$\tilde u_l = u_l - \uE[u_l\,|\, \bz]$,
and $\ba_l \tilde u_l$ is a linear estimator for
$\tilde \bu_{-l}$.
Using the orthogonality principle \cite{ChoiJBook2},
it can be shown that
\begin{align}
C_l 
= \beta_l - \frac{||\br_l ||^2}{\nu_l},
\end{align}
where
\begin{align}
\beta_l & = \uE[||\tilde \bu_{-l}||^2\,\bigl|\, \bz] 
= {\rm Tr}({\rm Cov}(\bu_{-l}\,|\, \bz)) \cr
\nu_l & = \uE[\tilde u_l^2 \,\bigl|\, \bz]
= {\rm Var}(u_l\,|\, \bz) \cr
\br_l & = \uE[\tilde \bu_{-l} \tilde u_l\,\bigl|\, \bz].
	\label{EQ:t0}
\end{align}
Thus, if the second order statistics are known,
it is possible to find $C_l$ 
and also $l^*$,
i.e., the node 
to be requested to send its measurement in each round
can be found with known second order statistics.

\section{Gaussian DAS}	\label{S:GDAS}

In this section, we assume that
$\{x_1, \ldots, x_K\}$ is a set of correlated
Gaussian random variables, i.e., $\bx$
follows a Gaussian distribution as follows:
\be
\bx \sim \cN(\bar \bx, \bR),
\ee
where $\bar \bx = \uE[\bx]$ and $\bR = {\rm Cov}(\bx)$.
In this case, we can have a closed-form expression for $C_l$
from $\bR$.

Let $\bar \bu = \uE[\bu]$,
$\bar u_l = \uE[u_l]$,
and $\bar \bz = \uE[\bz]$, which can be obtained
from $\bar \bx$ as they are subvectors of $\bar \bx$.
From \cite{Anderson79},
the conditional 
mean vector and covariance matrix of $\bu$ are 
\begin{align}
\uE[\bu \,|\, \bz] & = \bar \bu
+ \bR_{\bu, \bz} \bR_\bz^{-1} 
(\bz - \bar \bz) \cr
{\rm Cov}(\bu\,|\, \bz) & = \bR_\bu - 
\bR_{\bu, \bz} \bR_\bz^{-1} \bR_{\bu, \bz}^\rT ,
	\label{EQ:t1}
\end{align}
where $\bR_\bu = {\rm Cov}(\bu)$,
$\bR_{\bu, \bz} = \uE[(\bu - \bar \bu) (\bz - \bar \bz)^\rT]$,
and $\bR_\bz = {\rm Cov} (\bz)$.
Then,
all the terms in \eqref{EQ:t0} can be found
from the conditional covariance matrix of $\bu$,
i.e., ${\rm Cov}(\bu\,|\, \bz)$,
and the estimate of $\bu_{-l}$
that minimizes the MSE in \eqref{EQ:C_l} is given by
\be
\hat \bu_{-l}
= \uE[\bu_{-l} \,|\, \bz] + 
\br_l
\frac{u_l - \uE[u_l\,|\, \bz]}{{\rm Var}(u_l\,|\, \bz)},
\ee
where $\uE[\bu_{-l} \,|\, \bz]$ is the subvector
of $\uE[\bu \,|\, \bz]$ associated with $\bu_{-l}$, and
$\uE[u_l \,|\, \bz]$ is the $l$th element of 
$\uE[\bu \,|\, \bz]$, and 
${\rm Var}(u_l \,|\, \bz)$ is the $(l,l)$th element of 
${\rm Cov}(\bu \,|\, \bz)$.

It is noteworthy that
since $\beta_l$, $\br_l$, and $\nu_l$
are independent of $\bz$,
the optimal polling sequence 
(or the optimal order of nodes to transmit
their measurements) 
can be decided in advance.
That is, if $\bar \bx$ and $\bR$ are given
(for Gaussian $\bx$),
the BS is able to pre-determine\footnote{As a result,
Gaussian DAS is not exactly DAS as the available
data set, i.e., $\bz_t$, is not utilized to select
the next node for sensing/uploading. However,
as will be discussed in Section~\ref{S:MRA},
in Gaussian DAS, the node selection becomes dependent
on $\bz_t$ when 
there are some nodes that are unable to send their measurements
up on request.}
the order of the nodes in advance
to upload their measurements to minimize the MSE
in each round.
Note that this differs from the case with
$\bx$ that has a sparse representation
(not Gaussian) in \cite{Choi19}
(where the next node to transmit 
its measurement for DAS depends on $\bz_t$).

\begin{figure}[thb]
\begin{center}
\includegraphics[width=\figwidth]{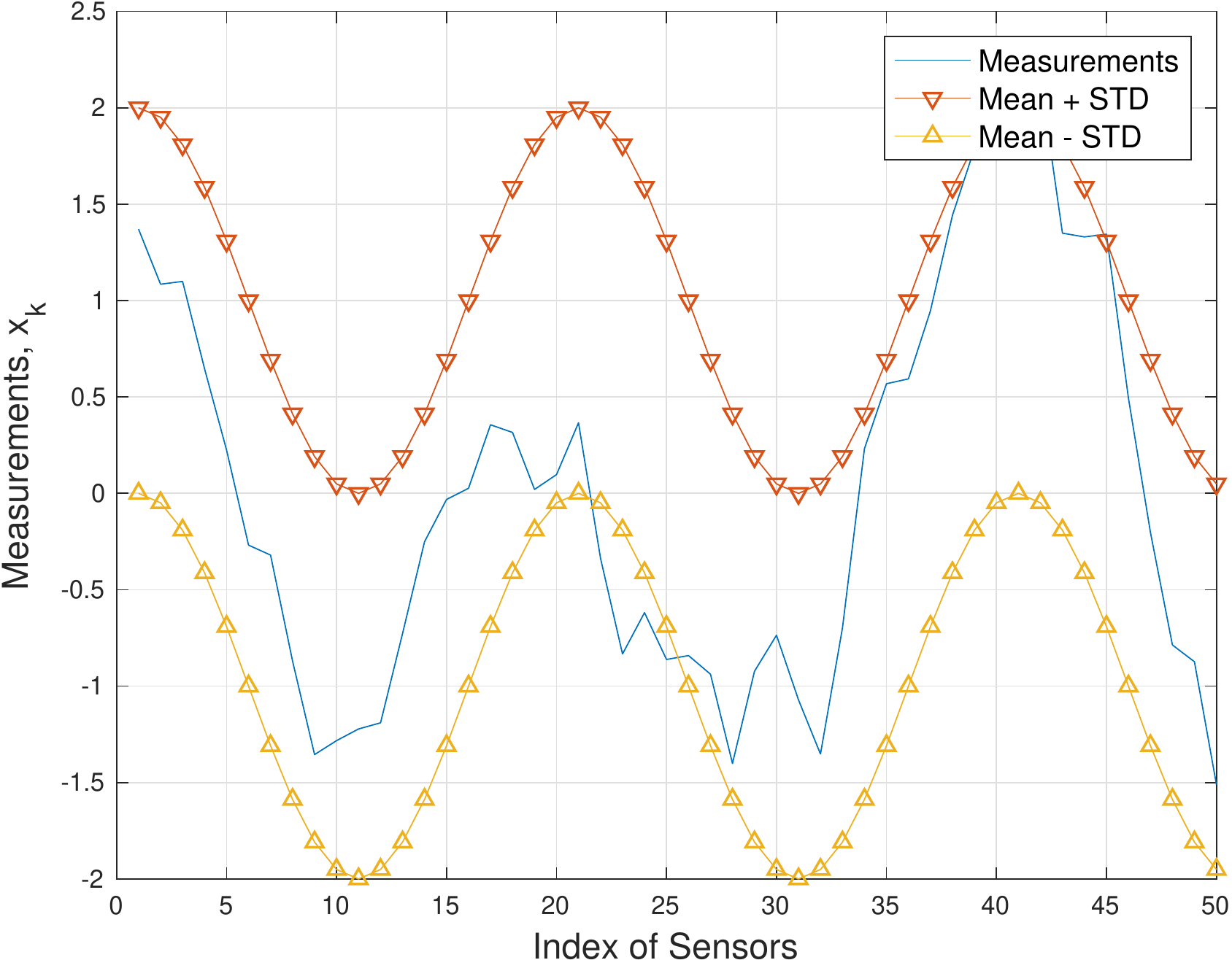} \\
(a) \\
\includegraphics[width=\figwidth]{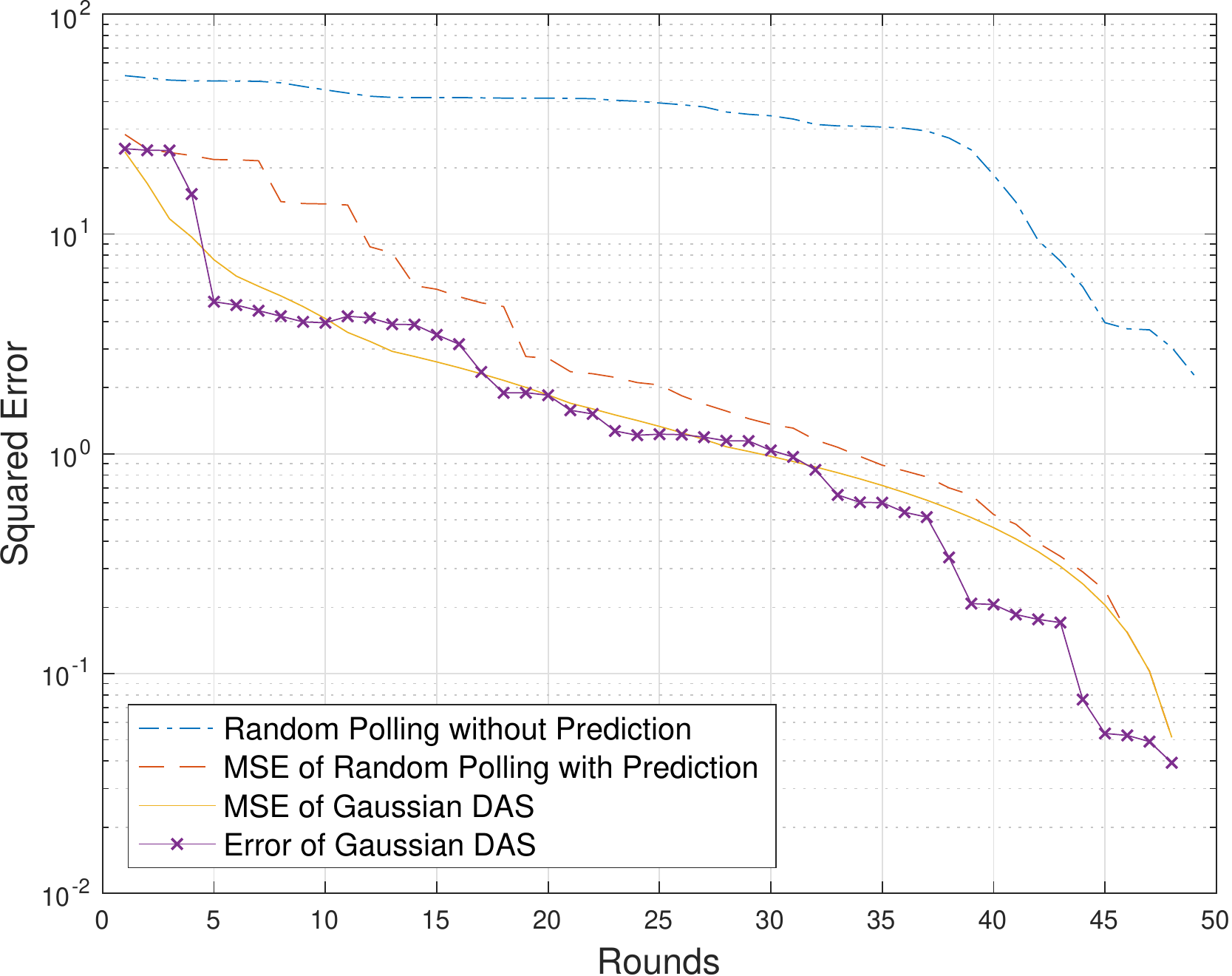} \\
(b) \\
\end{center}
\caption{Performance of data collection from
$K$ nodes:
(a) a realization of $\bx$ with the mean $\pm$ standard
deviation; 
(b) the MSE of Gaussian DAS and actual squared error in each
round.}
        \label{Fig:exam1}
\end{figure}

In Fig.~\ref{Fig:exam1},
we show the MSE of Gaussian DAS when the mean
of $\bx$ is given by
\be
\bar x_k = \cos \left( \frac{\pi}{5} (k -1) \right), 
\ k = 1,\ldots, K,
	\label{EQ:mean}
\ee
and the covariance matrix is 
\be
[\bR]_{k,k^\prime} = \rho^{|k - k^\prime|}, \ k,k^\prime 
\in \{1, \ldots, K\},
	\label{EQ:cov}
\ee
where $\rho = 0.95$.
A realization of $\bx$ is shown in Fig.~\ref{Fig:exam1} (a)
with the mean $\pm$ standard deviation (STD).
Fig.~\ref{Fig:exam1} (b)
shows the MSE in each round when Gaussian DAS
is used (by the solid line) and the actual squared error,
$||\bu_{-l} - \hat \bu_{-l}||^2$ is also shown (by the solid
line with cross markers). There are two additional
curves in 
Fig.~\ref{Fig:exam1} (b): one is the squared error 
without any prediction for $\bu_{-l}$, i.e.,
$||\bu_{-l}||^2$, with random polling
(by the dash-dotted line) and the MSE with prediction
for $\bu$ (by the dashed line).
Clearly, we can see that Gaussian DAS can help improve
the estimate of $\bx$ with a smaller number of measurements.

In practice, 
there can be nodes that do not have
measurements yet when requested. 
In addition, some nodes experiencing
deep fading cannot transmit their measurements
reliably. We discuss these issues in the next section
in order to justify the use of random access
for uploading measurements.

\section{Multichannel Random Access over Fading Channels}	\label{S:MRA}

In DAS, a node needs 
to have its data to upload when the BS requests.
However, the node may not have
its measurement as sensing is not carried out on time
or its transmit power may not be sufficiently high
to overcome fading. In this case,
the BS cannot receive any measurement. To mitigate this problem,
in this section, we consider 
multiple parallel channels and 
a different approach based on random access.

\subsection{Sequential Polling}

Suppose that node $k$ is asked by the BS to transmit its measurement
at round $t$.
Let $h_{k,t}$ denote the channel coefficient
to the BS from node $k$ 
at round $t$.
We assume time division duplexing (TDD) mode so that
node $k$ can estimate the channel coefficient from the request
signal (that includes a pilot sequence) from the BS.

Suppose that the received signal at the BS during
round $t$ is given by
\be
\br_t = h_{k,t} \sqrt{P_{k,t}} \bs_k + \bn_t,
\ee
where $P_{k,t}$ is the transmit power, $\bs_k$ is the encoded
signal vector to transmit measurement $x_k$, and
$\bn_t \sim \cC \cN(0, N_0 \bI)$ is the 
background noise.
The receive signal-to-noise ratio (SNR) becomes
\be
\gamma_k = \frac{|h_{k,t}|^2 P_{k,t}}{N_0}.
\ee
Suppose that the receive SNR has to be greater than
or equal to $\Gamma$ for successful decoding at the BS.
When node $k$ has a maximum transmit power, denoted by $P_{\rm max}$,
it cannot achieve the 
threshold SNR, $\Gamma$, if $\frac{|h_{k,t}|^2 P_{\rm max}}{N_0} < \Gamma$
(in this case, node $k$ decides not to upload its measurement).
In addition, node $k$ may not perform sensing yet.
Thus, the probability that node $k$ can transmit its measurement
is given by
\begin{align}
p_k 
& =  \Pr
\left(\left\{\frac{|h_{k,t}|^2 P_{\rm max}}{N_0} \ge \Gamma\right\}\cap
\{\mbox{$x_k$ is available} \}
\right) \cr
& = 
\Pr\left(\frac{a_{k,t} P_{\rm max}}{N_0} \ge \Gamma \right) W_k, 
	\label{EQ:p_k}
\end{align}
where $W_k = \Pr(\mbox{$x_k$ is available} )$.
In \eqref{EQ:p_k}, we assume that the channel 
coefficient is independent of the measurement availability at node $k$.
For convenience,
$p_k$ is referred to as the uploading probability.
For independent Rayleigh fading channels \cite{BiglieriBook}
\cite{ChoiJBook2},
it can be shown that
\be
f(a_{k,t}) = \frac{e^{- \frac{a_{k,t}}{\sigma_k^2}}}
{\sigma_k^2},
\ee
where 
$a_{k,t} = |h_{k,t}|^2$
and $\sigma_k^2 = \uE[a_{k,t}] = \uE[|h_{k,t}|^2]$\footnote{In general,
the channel 
coefficient can be expressed as $h_{k,t} = \sqrt{\sigma_k^2} \nu_{k,t}$,
where $\sqrt{\sigma_k^2}$ and and $\nu_{k,t}$
represent the long-term and short-term fading coefficients, 
respectively. Here, $\sigma_k^2$ is usually dependent on
the distance between node $k$ and the BS.}
We also assume that $h_{k,t}$ is independent in each round $t$.
It can be shown that
\begin{align}
p_k = 
\Pr\left(\frac{a_{k,t} P_{\rm max}}{N_0} \ge \Gamma \right) W_k 
= 
\exp\left(- \frac{\Gamma}{\Gamma_k}\right) W_k,
\end{align}
where $\Gamma_k = \frac{\sigma_k^2 P_{\rm max}}{N_0}$,
which is seen as the (maximum) average receive SNR.
Thus, if $\Gamma_k = \Gamma$ for all $k$ or node $k$'s 
$P_{\rm max}$ is adjusted to be proportional to $\Gamma_k$,
we have $p_k = e^{-1} W_k \le e^{-1}$,
which means that the uploading probability
might be low and less than $e^{-1}$
(provided that the average receive SNR is set to the threshold SNR).

If the BS performs sequential
polling for DAS, where in each round only one node is selected,
after $K$ rounds, the average number of measurements
becomes $\sum_{k=1}^K p_k \le K$
(i.e., due to fading and unavailability of measurement,
the BS may not able to collect all the measurements
after $K$ rounds).
Thus, more than $K$ rounds would be required to obtain
$\bx$ or a good estimate of $\bx$.

To shorten the time to obtain $\bx$ or a good estimate of $\bx$,
multiple parallel (orthogonal) channels can be considered.
Suppose that there are $N$ parallel multiple access 
channels. In this case,
in each round, the BS can send request signals to $N$
nodes\footnote{In this case, each requested node
can transmit its measurement through a dedicated channel
without any interference thanks to $N$ parallel channels.} 
and the total number of rounds can be reduced 
by a factor of $N$.
In this case, on average, we also have 
$\sum_{k=1}^K p_k$ measurements after $\frac{K}{N}$ rounds.
In addition, in each round, 
the average number of nodes that
can upload measurements is 
\be
N_{\rm sp} = N p,
	\label{EQ:Nsp}
\ee
if $p_k = p$ for all $k$.
In addition, at round $t$,
the number of the nodes that 
successfully transmit their measurements,
which is the length of $\bz$ at round $t$,
is a random variable. 
For convenience, let $K_t$ denote
the length of $\bz$ at round $t$.
Then, $K_t - K_{t-1}$ is a binomial 
random variable with parameter $N$ and $p$
if there are more than $N$ nodes with the measurements
that are not uploaded yet (at round $t-1$).

There are few remarks.
\begin{itemize}
\item
Note that when the channel coefficient, $h_{k,t}$,
varies from a round to another,
the BS can transmit again a request signal to a node
that had been requested, but was unable to transmit
its measurement (e.g., due to deep fading) in a past round.
In this case, the maximum number of rounds can be greater than 
$\frac{K}{N}$.

\item The nodes that can transmit their measurements
up to round $t$ cannot be pre-determined 
because some nodes are unable to transmit. 
As a result, the polling order for Gaussian DAS is not
pre-determined and has to be
adaptively decided (as expected in DAS), 
which differs from the case
discussed in Section~\ref{S:GDAS} (where it is assumed
that $x_k$ is always available when node $k$ is requested to send
its measurement).
\end{itemize}

\subsection{Multichannel ALOHA}

Suppose that the uploading probability, $p_k$, is low,
which may result in most $N$ channels not being used.
To mitigate this problem,
we can use random access rather than polling.
In particular, we consider multichannel ALOHA
\cite{Shen03} \cite{Chang15}
with $N$ parallel channels.
We assume that the BS chooses $Q$ nodes to request to upload 
their measurements in each round with no dedicated
channel for each user (in this case, $Q$ can be greater than $N$).

When multichannel ALOHA  is used,
a channel can be chosen by multiple nodes,
which results in packet collision (which is not the
case when polling is used). 
Note that
the capture effect \cite{Goodman87} allows 
the BS to decode the strongest signal(s) when there is packet
collision. However, for simplicity,
in this paper, it is assumed
that no packet is decodable when there is packet collision. 
To see the performance, let $p_k = p$ for all $k$ for simplicity. 
Then, the average number of nodes that can 
successfully
transmit their measurements 
in each round becomes
\begin{align}
N_{\rm ra}
& = \sum_{q=1}^Q q \left(1 - \frac{1}{N} \right)^{q-1} \Pr(q; Q) \cr
& = \sum_{q=1}^Q q \left(1 - \frac{1}{N} \right)^{q-1} 
\binom{Q}{q} p^q (1-p)^{Q-q} \cr
& = Q p \left( 1 - \frac{p}{N} \right)^{Q-1},
	\label{EQ:Nra}
\end{align}
where $\Pr(q; Q)$ represents
the probability that the number of nodes that can send and upload
is $q$.
For a sufficiently large $N$, we can see that
$Q = \frac{N}{p}$ can maximize
$N_{\rm ra}$.
In this case, we have 
$$
N_{\rm ra} = N e^{-1}.
$$
Therefore, compared with $N_{\rm sp}$ in \eqref{EQ:Nsp},
it can be seen 
that multichannel ALOHA
can provide a better performance
(in terms of the average number of nodes
that can successfully sense and upload measurements per round) 
than sequential polling when 
\be
p < e^{-1} \approx 0.3679,
	\label{EQ:pe0}
\ee
although there are packet collisions.
As mentioned earlier, the uploading probability
under independent Rayleigh fading 
becomes less than $e^{-1}$
(provided that the average receive SNR
is set to the threshold SNR),
which demonstrates that 
\eqref{EQ:pe0} can hold with
a limited transmit power at nodes under Rayleigh fading.

In Gaussian DAS with multichannel ALOHA, 
we assume that the nodes that can successfully transmit
their measurements (without collisions) are not asked
again to transmit their measurements.
Let $\cD (t)$ denote the index set of the nodes
that send their measurements to the BS 
at round $t$ without collisions,
while $\cB (t)$ denotes the 
index set of the nodes that are requested by the BS to transmit
their measurements. In multichannel ALOHA,
to maximize the number of the measurements without collisions,
we need to set the size of $\cB(t)$ to $Q$, i.e.,
$|\cB(t)| = Q$.
Note that $\cD(t)$ is a set of random indices
(due to collisions) and $\cD(t) \subseteq \cB(t)$.
For convenience, let
$$
\cA (t) = \cup_{i=0}^{t} \cD(i) = \cA(t-1) \cup \cD(t),
$$
where 
$\cA(-1) = \emptyset$ and
it can be shown that $\cA(t-1) \cap \cD(t) = \emptyset$.
Then, the elements of $\bz$ are the elements
of $\bx$ corresponding to $\cA (t-1)$ and $Z_t = |\cA(t-1)|$
which is the length of $\bz_t$ at round $t$.

Denote by $\bx_{\cD(t)}$ 
and $\bx_{\cA(t)}$ the subvectors
of $\bx$ according to $\cD(t)$ and $\cA (t)$, respectively.
Then, we have 
\begin{align}
\bz_t = \bx_{\cA(t-1)} \ \mbox{and} \
\bu_t = \bx_{\cA(t-1)^c}.
\end{align}
Based on Gaussian DAS, the BS chooses $Q$ nodes\footnote{If 
the number of the nodes that do not transmit their measurements yet
is less than $Q$, i.e., $K - K_t < Q$, $Q$ becomes $K- K_t$.}
from $\bu$ in round $t$ that minimize the conditional MSE and
send request signals to them. Among $Q$ nodes, there might be
a fraction of them that are able to sense
and upload their measurements
(without collisions and deep fading).
Their measurements are to form $\bx_{\cD(t)}$.

In summary, the following pseudo-code 
is presented for Gaussian DAS when $\bx$
is to be estimated with measurements from $\bar K$ nodes,
where $\bar K \le K$.

\begin{itemize}
\item[S0)] Set $t = 0$ and $\cA(-1) = \emptyset$. 
The BS sends request
signals to randomly selected $Q$ nodes from $K$ nodes,
where $K \gg Q$.
\item[S1)] The requested nodes 
upload their measurements if they can (i.e., if they
have measurements and do not have deep fading) according to 
the uploading conditions in the probability in \eqref{EQ:p_k}.
\item[S2)] The BS receives the measurements
from nodes associated with $\cD(t)$ and update
$\cA(t) = \cA(t-1) \cup \cD(t)$.
If $|\cA(t)| \ge \bar K$, the BS stops collecting measurements.
\item[S3)] Let $t \leftarrow t +1$. The BS performs Gaussian DAS to choose
the next $Q$ nodes
that minimize the conditional MSE (for given
$\bz_t = \bx_{\cA(t-1)}$), and sends request signals to them.
\item[S4)] Move to S1).
\end{itemize}

The minimum number of rounds that the BS has 
$\bar K$ measurements is random, 
which is given by
\be
T_{\rm ma} (\bar K) = \min\left\{T\,:\,  \bar K \le \sum_{t = 0}^{T-1} 
|\cD(t)|\right\}.
	\label{EQ:Tma}
\ee
If $\bar K \le K - Q$,
$|\cD(t)|$ can be seen as 
independent identically distributed (iid) random variables
with the mean in \eqref{EQ:Nra}.
Using Wald's identity 
\cite{Mitzenmacher},
from \eqref{EQ:Tma},
the average number of rounds to get $\bar K$ measurements is given by
\begin{align}
\bar T_{\rm ma} (\bar K)
= \uE[T_{\rm ma} (\bar K)] 
\ge \frac{\bar K}{Q p 
\left(1 - \frac{p}{N} \right)^{Q-1}}
\approx \frac{\bar K}{N e^{-1}},
	\label{EQ:T1}
\end{align}
with $Q = \frac{N}{p}$ (if $Q \le K$).

When sequential polling is used, 
the average number of rounds to get $\bar K$ measurements becomes
\be
\bar T_{\rm sp} (\bar K) \ge \frac{\bar K}{N p}.
	\label{EQ:T2}
\ee
From \eqref{EQ:T1} and \eqref{EQ:T2},
for a low $p$,
the number of rounds to get $\bar K$ measurements
with multichannel ALOHA
can be much smaller than that with
sequential polling.

\section{Gaussian DAS with Model Selection
as a Multi-armed Bandit} \label{S:MAB}

Suppose that there are $M$ models,
where $M \ge 2$, for the measurements, $\bx$.
Each model has a different parameter set, i.e.,
$\{\bar \bx_m, \bR_m\}$,
where $\bar \bx_m$ and $\bR_m$
represent the mean vector and covariance matrix, respectively,
of $\bx$ under model $m$.
If the BS knows that the measurement
vector, $\bx$, follows a model, say model $m$, in advance,
it can set $\bar \bx = \bar \bx_m$
and $\bR = \bR_m$ and perform Gaussian DAS
to collect measurements from nodes.
However, if the BS does not know the model,
it is required to find the correct model when
nodes transmit their measurements (in this section,
we only consider multichannel ALOHA for DAS).
To this end, in this section, we consider a multi-armed bandit problem
for the model selection in conjunction
with Gaussian DAS.

Let $\bar m \in \{1, \ldots, M\}$ denote the correct model for
convenience.
Since there are $M$ models,
let $\cD(t; m)$ denote 
the index set of the nodes that transmit their measurements 
without collisions
in round $t$ under model $m$.
In addition, denote by $\cB(t;m)$ 
the index set of the nodes that are requested
to transmit their measurements at round $t$ under model $m$.
If the model chosen at round $t$ is denoted by
$m(t)$, we have
\be
\cA (t) = \cup_{i=0}^t \cD(i; m (i)),
\ee
which depends on $\{m(0), \ldots, m(t)\}$,
while $\cD (t;m (t)) \subseteq \cB(t; m(t))$.
As mentioned earlier, due to fading and availability
of measurement at each node, some nodes
are not able to upload their measurements although
they are requested to transmit.
In addition, the BS may not be able to receive
measurements from some nodes responding
and uploading, but experiencing collisions
in multichannel ALOHA.
As a result, 
a fraction of the nodes in $\cB(t; m(t))$
can succeed to transmit their measurements,
which are associated with $\cD(t; m(t))$.

At round $t$, $\bz_t = \bx_{\cA(t-1)}$
is given regardless of the current model that the BS is to choose.
If the BS chooses model $m$, it receives
$\bx_{\cD(t;m)}$.
The cost of choosing model $m$ after round $t$
can be given as
\be
Y_{t;m} = 
\frac{||\bx_{\cD(t;m)} - \hat \bx_{\cD(t;m)} ||^2}{
\uE_m[||\bx_{\cD(t;m)} - \hat \bx_{\cD(t;m)} ||^2]},
	\label{EQ:Ytm}
\ee
where 
$\uE_m [\cdot]$ represents the expectation under model $m$
and
$\hat \bx_{\cD(t;m)}$ is the (conditional) MMSE estimate
of $\bx_{\cD(t;m)}$ for given $\bz_t$ under model $m$. That is,
$\hat \bx_{\cD(t;m)}  = \uE_m [\bx_{\cD(t;m)} \,|\, \bz_t]$.
If model $m$ is correct, i.e., $m = \bar m$, it is expected that
\be
\uE_{\bar m} [Y_{t;m}] = 1.
	\label{EQ:Yc}
\ee
However, if model $m$ is incorrect, 
i.e., $m \ne \bar m$, we have
\be
\uE_{\bar m} [Y_{t;m}] > 1.
	\label{EQ:Yi}
\ee
If the BS can choose the correct model, i.e., 
model $\bar m$, it can not only have
a low MSE, but also achieve a good estimate
of $\bx$ with a small number of 
the nodes that transmit measurements thanks to
DAS.

The model section problem with Gaussian DAS
can be seen as a multi-armed bandit problem as 
the BS can choose a set of nodes 
(e.g., $Q$ nodes with multichannel ALOHA) in each round
according to a selected model $m$.
The BS needs to have both exploration and exploitation
\cite{Sutton98}, because it does not know the correct model
in advance, but needs to decide a model 
while performing DAS.
There are a number of different 
multi-armed bandit 
algorithms \cite{Vermorel05} \cite{Scott10} \cite{Bubeck12}.
However, since we are interested in applying
a multi-armed bandit formulation to
the model selection with DAS,
we only consider one algorithm,
which is called the ``softmax" algorithm
\cite{Luce59} \cite{Sutton98}.

Let 
\be
\psi_{t;m} = \frac{
\sum_{i = 0}^t Y_{i;m(i)}\indicator(m(i) = m) }{
\sum_{i = 0}^t \indicator(m(i) = m)},
	\label{EQ:psi}
\ee
which is seen as the sample mean of the cost of model $m$
up to time $t$.
In \eqref{EQ:psi}, $\indicator(A)$
is the indicator function that becomes 1 if event
$A$ is true and 0 otherwise.
At round $t+1$, the BS chooses a model according to
the following probabilities:
\be
P_m (t+1) = 
\frac{e^{-\frac{\psi_{t;m}}{\tau}}}
{\sum_{m=1}^M e^{-\frac{\psi_{t;m}}{\tau}}},
	\label{EQ:Pm}
\ee
where $\tau$ is a temperature parameter
that controls the randomness of the choice
(or is used to enjoy the trade-off between exploration
and exploitation).
For example, if $\tau \to \infty$, $P_m (t+1) \to \frac{1}{M}$,
which means the probability that model $m$ is chosen
is the same for all $m$. 
With a finite $\tau$,
from \eqref{EQ:Pm}, we expect that
a model associated with a lower cost
has a higher probability to be selected.
From \eqref{EQ:Yc} and \eqref{EQ:Yi},
the mean of the cost of the correct
model, i.e., $Y_{t; \bar m}$ is smaller 
than that of incorrect ones. Thus, it is expected that
the probability that the correct model
is chosen becomes the highest as $t$ increases.
Note that in order to have valid sample means,
at the initial exploration stage, we need 
to select all the models at least once.
Thus, we assume that
$$
m(t) = m+1, \ t = 0, \ldots, M-1.
$$
Then, from round $M$, a model is to be chosen
according to \eqref{EQ:Pm}.

There are a few issues we do not address in this paper regarding
the model selection with DAS as follows.
\begin{itemize}
\item Performance analysis: The multi-armed bandit problem
with the cost function in \eqref{EQ:Ytm}
differs from a standard 
or conventional multi-armed bandit problem,
where the cost (or reward) is assumed
to be bounded and 
iid.
As $t$ increases, the length of $\bz_t$
grows and the prediction error
has different statistical properties 
although it is normalized by its mean as shown in 
\eqref{EQ:Ytm}. Thus, $Y_{t;m}$ may not be iid for a given $m$
(in addition, $Y_{t;m}$ depends on the past selected models,
i.e., $\{m(0), \ldots, m(t-1)\}$ through $\bz_{t}$ or $\cA(t-1)$
as mentioned earlier).
As a result, existing approaches for the performance
analysis (which are based on the assumption of bounded iid rewards)
cannot be used,
which means that a new tool is to be developed for
the performance analysis.

\item We assume that each model is specified by
its mean and covariance matrix
(i.e., $\{\bar \bx_m, \bR_m\}$, $m = 1,\ldots, M$). In practice,
it is necessary to find (or estimate) them.
To this end, the BS needs to collect a sufficient
number of measurements from nodes
without DAS and perform classifications
using any clustering algorithms \cite{Duda01}.
Alternatively, DAS may be combined into clustering,
which is a further research topic to be studied in the future.
\end{itemize}

\section{Simulation Results}	\label{S:Sim}

In this section, we present
simulation results for Gaussian DAS.
We first consider the performance with
sequential polling and
multichannel ALOHA
to upload measurements.
Then, simulation
results for the model selection in conjunction with Gaussian DAS
are presented when 
multichannel ALOHA is used.

\subsection{Sequential Polling versus Multichannel ALOHA}	\label{SS:SP}

In this subsection, we assume that
the mean and covariance are given as
in \eqref{EQ:mean} and \eqref{EQ:cov},
respectively.

\begin{figure}[thb]
\begin{center}
\includegraphics[width=\figwidth]{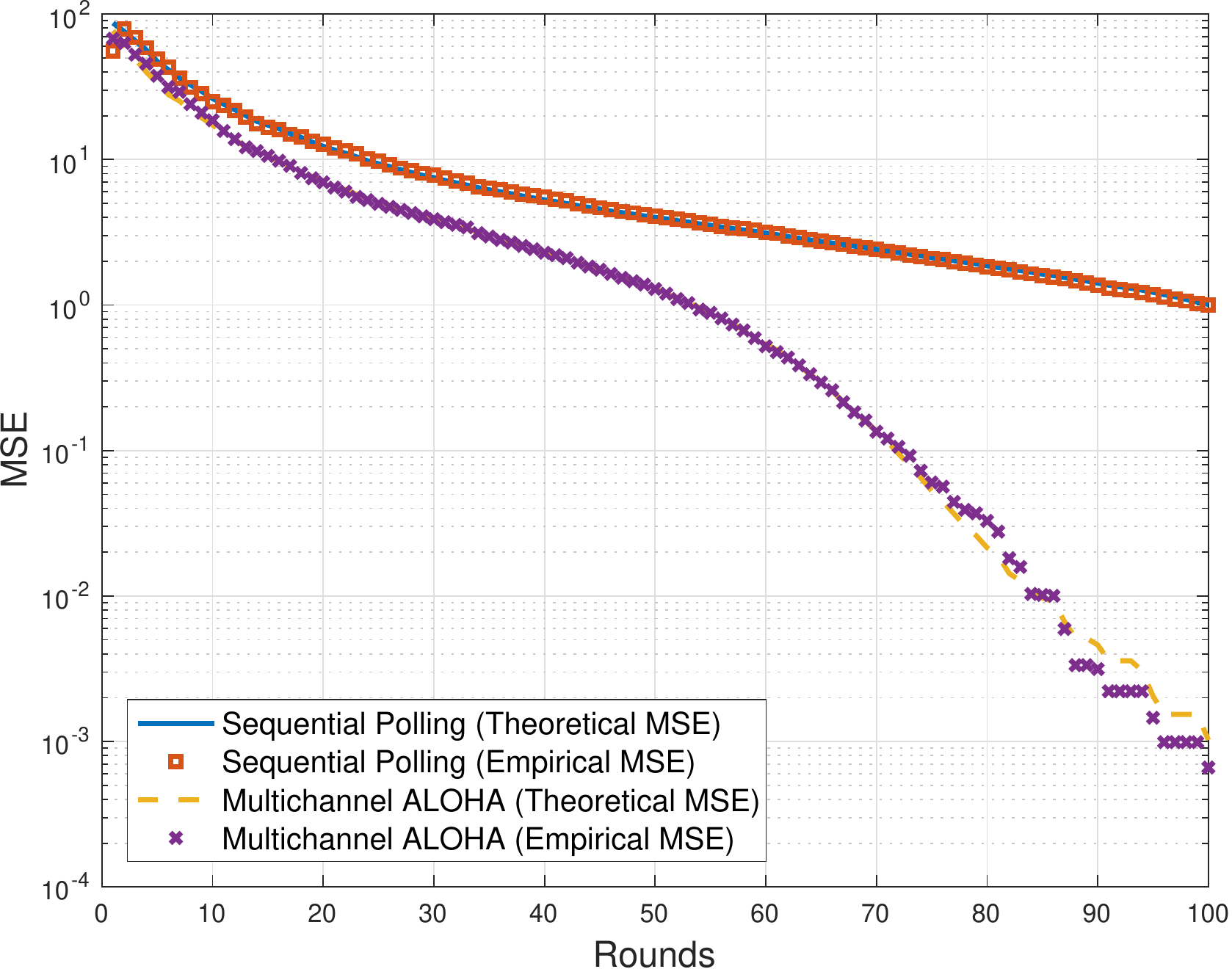} 
\end{center}
\caption{MSEs with sequential polling 
and multichannel ALOHA as functions of rounds 
when $K = 100$, $p = 0.2$, and $N = 4$.}
        \label{Fig:plt1}
\end{figure}

Fig.~\ref{Fig:plt1} shows 
the MSEs with sequential polling 
and multichannel ALOHA as
functions of rounds 
when $K = 100$, $p = 0.2$, and $N = 4$.
For multichannel ALOHA, $Q$ is set to 
$\min\{K - K_t, \frac{N}{p} \}$ in all simulations
in this section.
From \eqref{EQ:T1}
and \eqref{EQ:T2},
if $\bar K = 75$, we have
$$
T_{\rm ma} = 50.96 
\ \mbox{and} \ 
T_{\rm sp} = 93.75.
$$
That is, the BS is able to have 75\% 
of total measurements with about 51 rounds
with multichannel ALOHA and 94 rounds 
with sequential polling.
As shown in 
Fig.~\ref{Fig:plt1},
we can see that the MSE with multichannel
ALOHA at $t = 51$ is similar to that with 
sequential polling at $t = 94$.
This demonstrates that multichannel ALOHA
can provide more efficient uploading performance
than sequential polling in Gaussian DAS.

\begin{figure}[thb]
\begin{center}
\includegraphics[width=\figwidth]{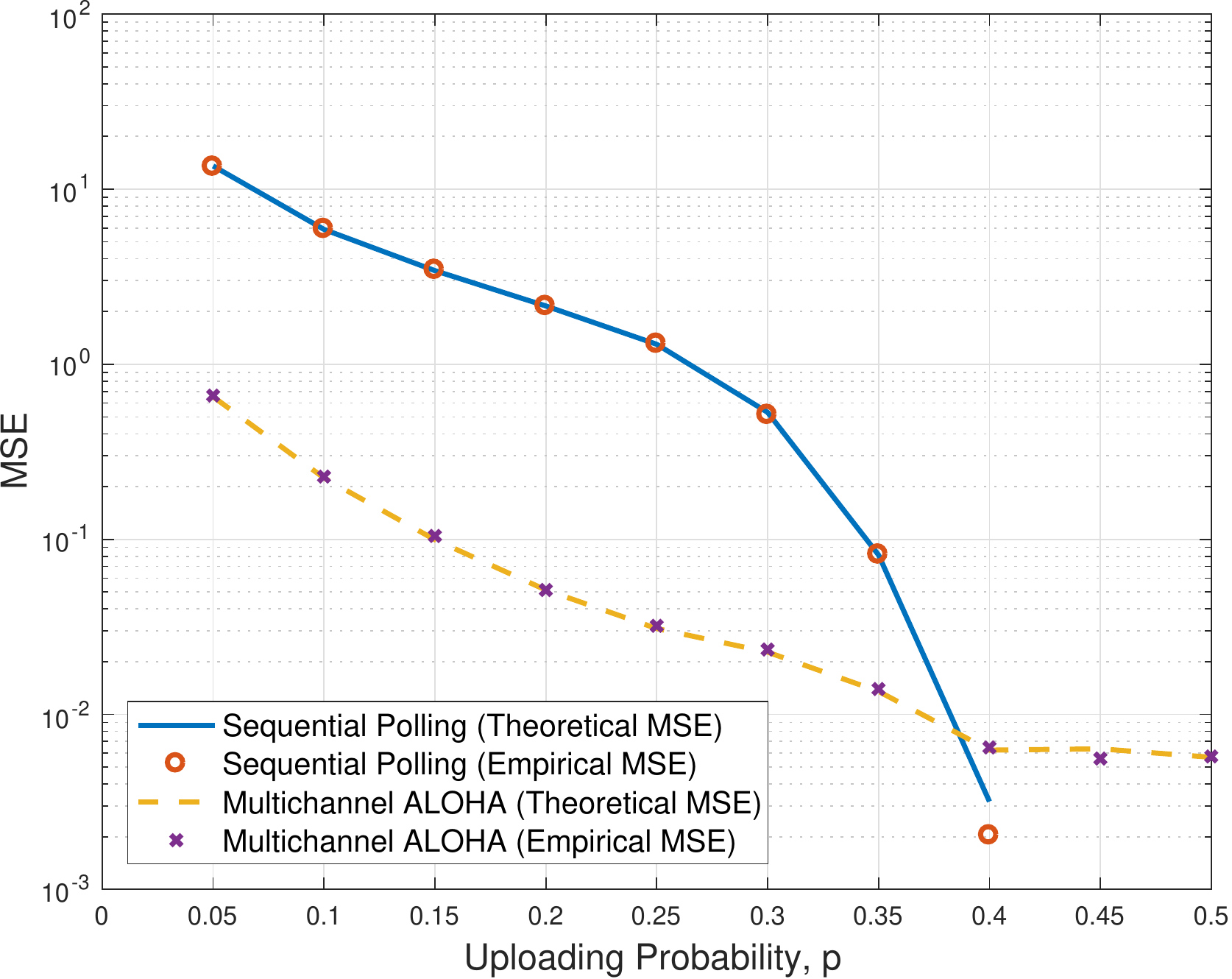} 
\end{center}
\caption{MSEs with sequential polling 
and multichannel ALOHA as functions of uploading probability, $p$,
after $T = 75$ rounds when $K = 100$ and $N = 4$.}
        \label{Fig:plt2}
\end{figure}

In order to see the impact of uploading probability,
$p$, on the performance, we run simulations
with different values of $p$ when
$K = 100$ and $N = 4$ and show the results after
$T = 75$ rounds in Fig.~\ref{Fig:plt2}.
As $p$ increases, the number of measurements
at the BS increases and a lower MSE is expected.
We can also confirm that if $p > e^{-1}$,
sequential polling performs
better than multichannel ALOHA in Fig.~\ref{Fig:plt2}.

\begin{figure}[thb]
\begin{center}
\includegraphics[width=\figwidth]{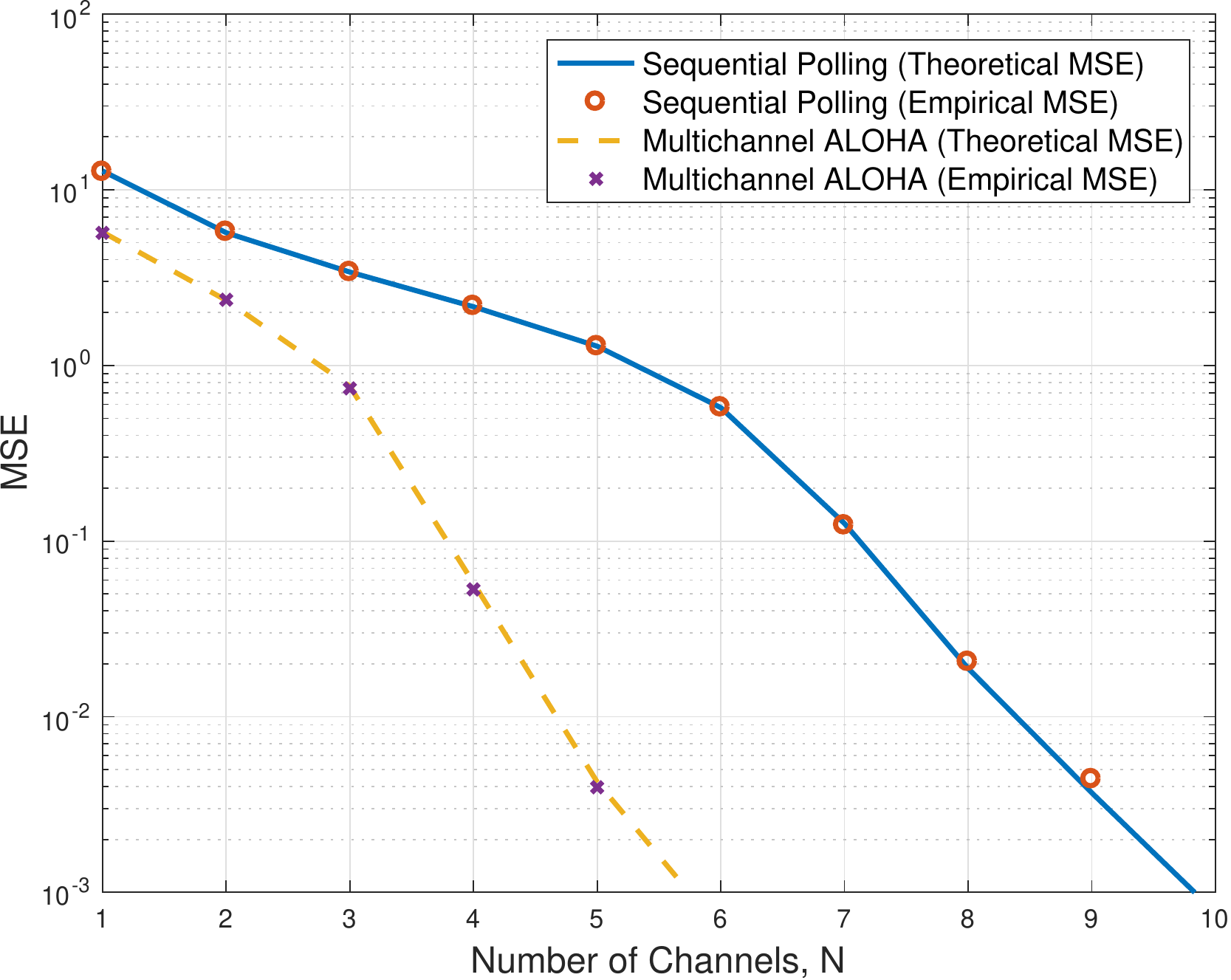} 
\end{center}
\caption{MSEs with sequential polling 
and multichannel ALOHA as functions of 
the number of channels, $N$,
after $T = 75$ rounds when $K = 100$ and $p = 0.2$.}
        \label{Fig:plt3}
\end{figure}

Fig.~\ref{Fig:plt3} shows the MSEs with sequential polling 
and multichannel ALOHA as functions of 
the number of channels, $N$,
after $T = 75$ rounds when $K = 100$ and $p = 0.2$.
As expected, the MSE decreases with $N$ as more
measurements are available.
We also see that multichannel ALOHA can provide
a better performance than sequential polling as $p = 0.2 < e^{-1}$.

In Figs.~\ref{Fig:plt1},~\ref{Fig:plt2}, and \ref{Fig:plt3},
the MSE (from \eqref{EQ:C_l})
is shown with the empirical MSE that
is obtained by taking an average of 100 runs,
where each run has an independent realization of $\bx$
for given $\bar \bx$ and $\bR$.
In general, we can see that the empirical MSE 
agrees with the MSE.
Note that the MSE in
Figs.~\ref{Fig:plt1},~\ref{Fig:plt2}, and \ref{Fig:plt3}
is also obtained by taking an average of 100 runs,
because the nodes
that send their measurements in each round is also random
(due to independent fading channels).

\subsection{Model Selection with Gaussian DAS}

In this subsection, we present simulation
results for the model selection with Gaussian DAS.
We assume that 
there are $M = 5$ different models.
Model 1 is the same one used in Subsection~\ref{SS:SP},
i.e., the mean vector and covariance matrix are
given as in \eqref{EQ:mean} and \eqref{EQ:cov},
respectively.
For the other 4 models, we have
\begin{align*}
[\bar \bx_2]_k & = \sin \left( \frac{\pi}{5} (k -1) \right), \ 
[\bar \bx_3]_k = -\cos \left( \frac{\pi}{5} (k -1) \right) \cr
[\bar \bx_4]_k & = -\sin \left( \frac{\pi}{5} (k -1) \right), \
[\bar \bx_5]_k = 0, \ k = 1,\ldots,K.
\end{align*}
For the covariance matrices,
we use the discrete cosine transform (DCT) of size
$K \times K$, 
which is unitary and denoted by $\bPsi$.
Let $\bpsi_k$ denote the $k$th column of $\bPsi$.
Then, we have
\begin{align*}
\bR_m = c_m \left(\sum_{k=m-1}^{J+m-2} \bpsi_k \bpsi_k^\rH + 0.1 \bI
\right),
\ m = 2, \ldots, M,
\end{align*}
where $J = 3$ and $c_m$ is a normalizing constant to make
${\rm Tr}(\bR_m) = K$ for $m \in \{2,\ldots,M\}$
(as that for $m = 1$).
To generate $\bPsi$ we use ``dctmtx" command in MATLAB.

\begin{figure}[thb]
\begin{center}
\includegraphics[width=\figwidth]{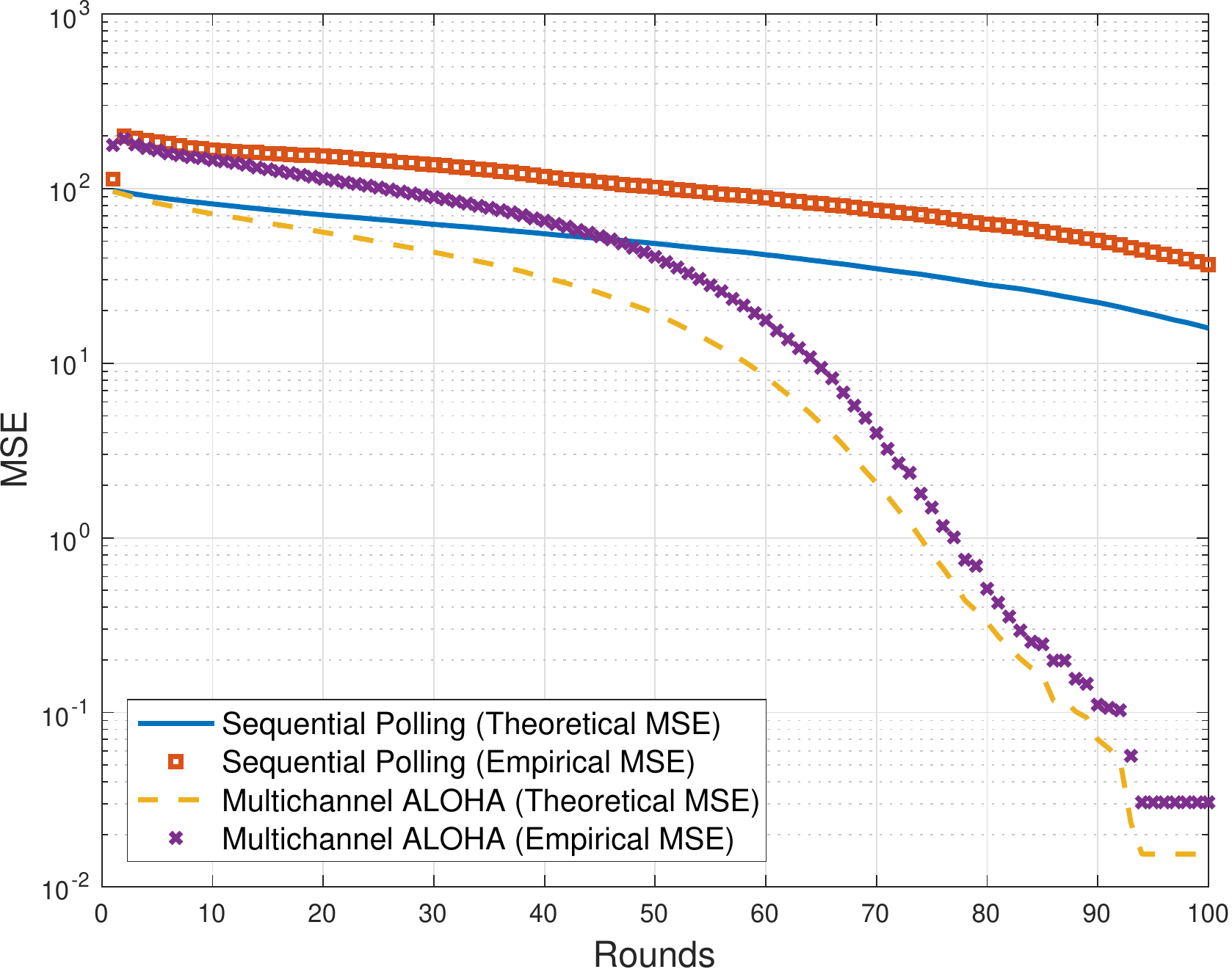} 
\end{center}
\caption{MSE with the correct model 
(i.e., model 1) and empirical MSE with an incorrect model (i.e., model 2)
as functions of rounds when $K = 100$, $p = 0.2$, and $N = 4$.}
        \label{Fig:w_plt1}
\end{figure}

In Fig.~\ref{Fig:w_plt1},
we show the (theoretical) MSE that is obtained with the correct model 
(i.e., model 1) and 
the empirical MSE that is obtained
with an incorrect model (i.e., model 2)
as functions of rounds when $K = 100$, $p = 0.2$, and $N = 4$. 
Since a wrong model is used in Gaussian DAS,
the empirical or sample MSE is higher
than the MSE as shown in 
Fig.~\ref{Fig:w_plt1}.

\begin{figure}[thb]
\begin{center}
\includegraphics[width=\figwidth]{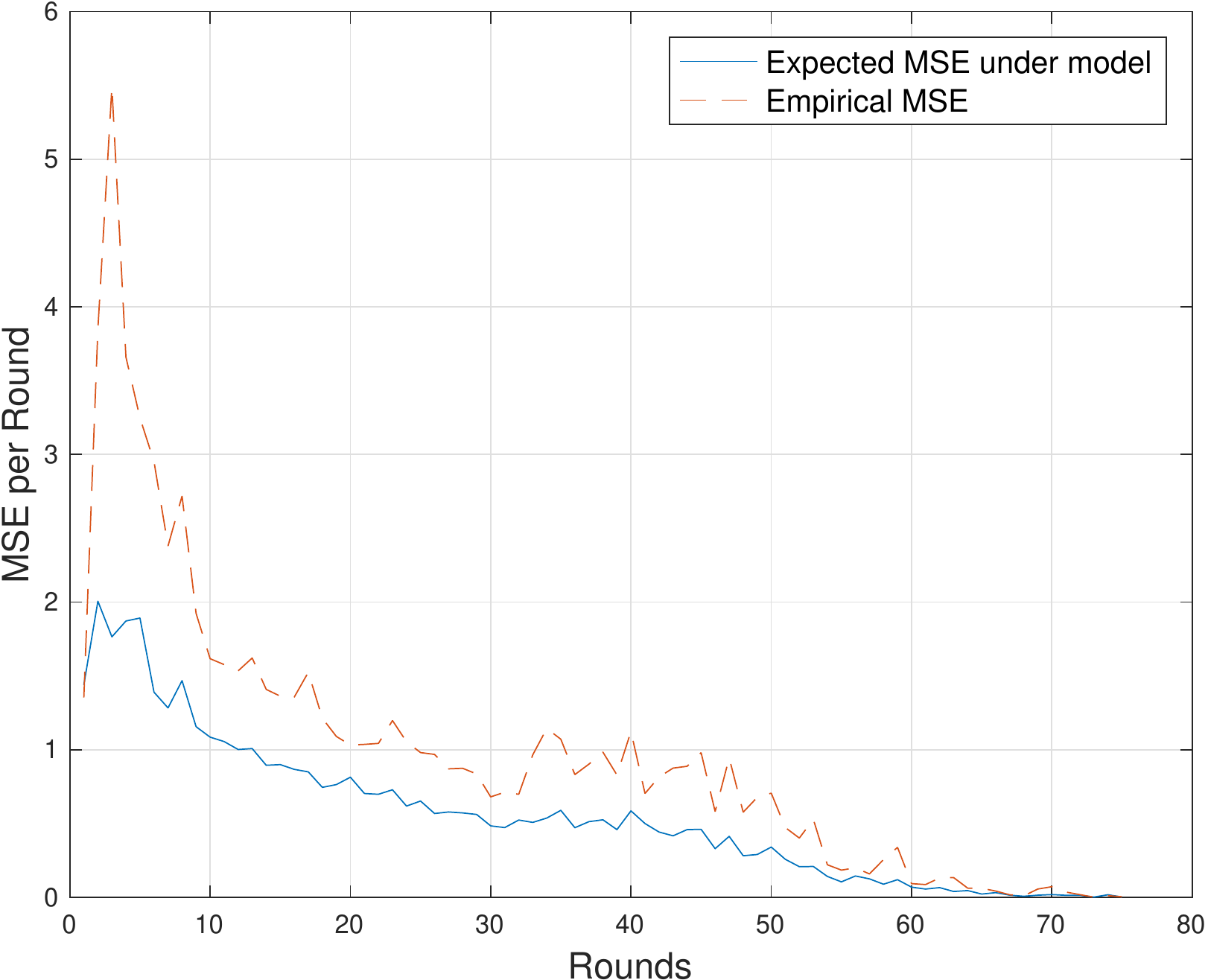} \\
(a) \\
\includegraphics[width=\figwidth]{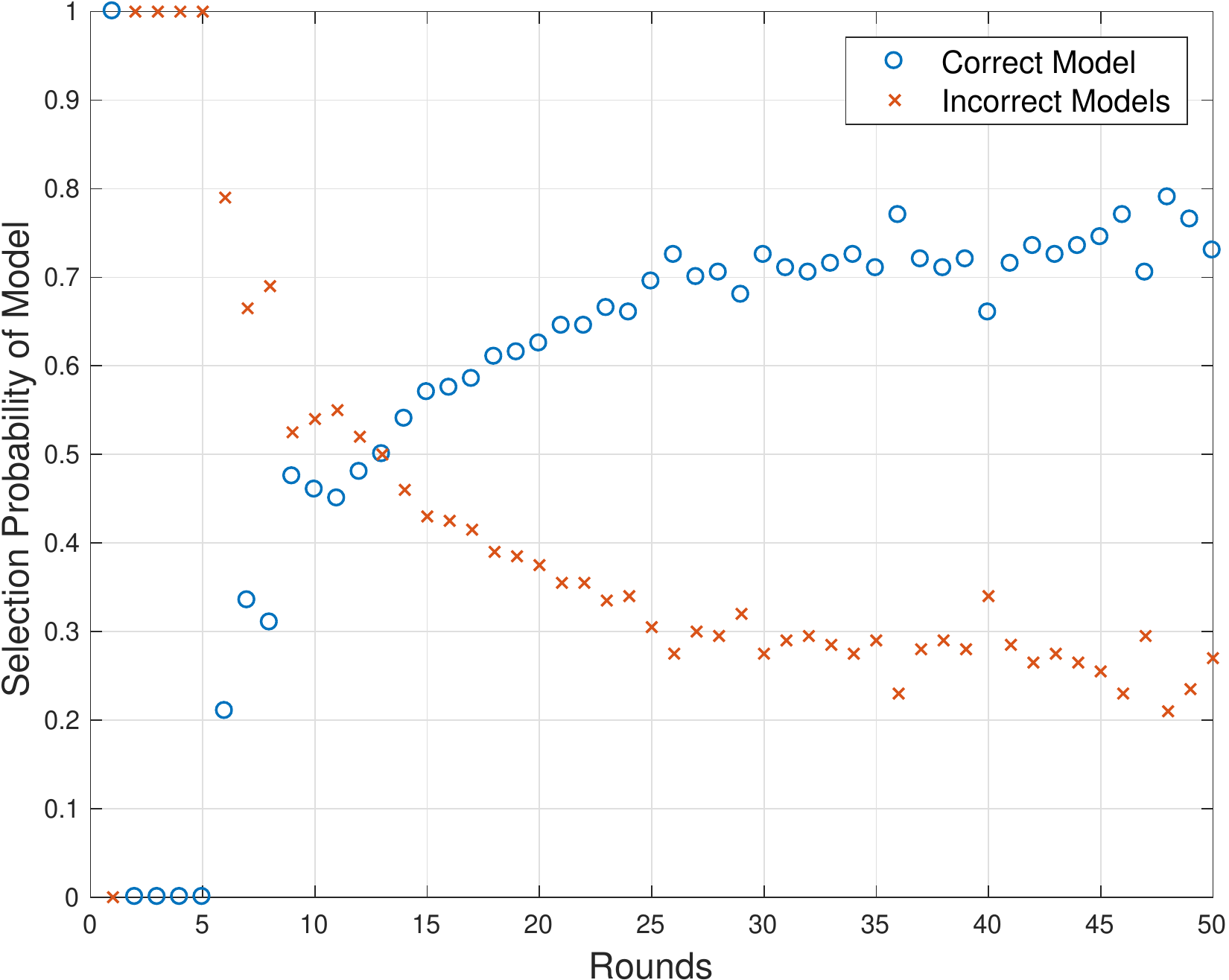} \\
(b) \\
\end{center}
\caption{Performance of model selection with Gaussian DAS
when $(K,p,N) = (100, 0.2, 4)$, $\tau = 1$
for softmax, and the correct
model is model 1: (a) MSE per round; (b) Selection probability of model
($\circ$ marks for the probability of selecting model 1
and $\times$ marks for the probability of selecting
incorrect models (i.e.,  $m \in \{2, \ldots, M\}$).}
        \label{Fig:x_plt1}
\end{figure}

We now only consider multichannel ALOHA to upload data from nodes.
In Fig.~\ref{Fig:x_plt1} (a),
we show the MSE per round,
i.e., $\uE[||\bx_{\cD(t)} - \hat \bx_{\cD(t)}||^2]$,
and its empirical one by taking average from independent 200 runs
when $(K,p,N) = (100, 0.2, 4)$, $\tau = 1$
for softmax, and the correct
model is model 1.
We can see that the MSE per round decreases
because of more measurements as well as a high
probability of selecting the correct model,
which is demonstrated in 
Fig.~\ref{Fig:x_plt1} (b).

\begin{figure}[thb]
\begin{center}
\includegraphics[width=\figwidth]{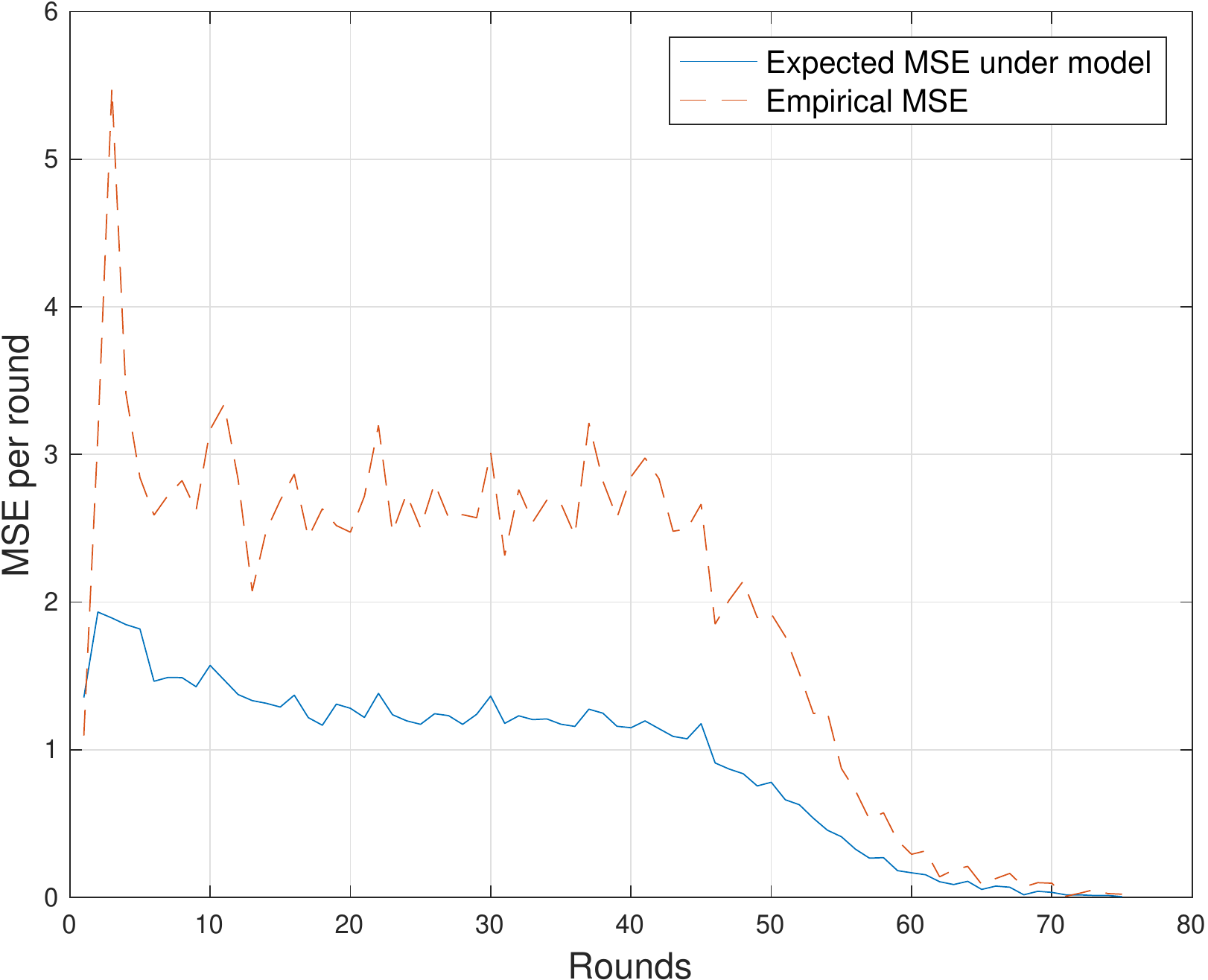} \\
(a) \\
\includegraphics[width=\figwidth]{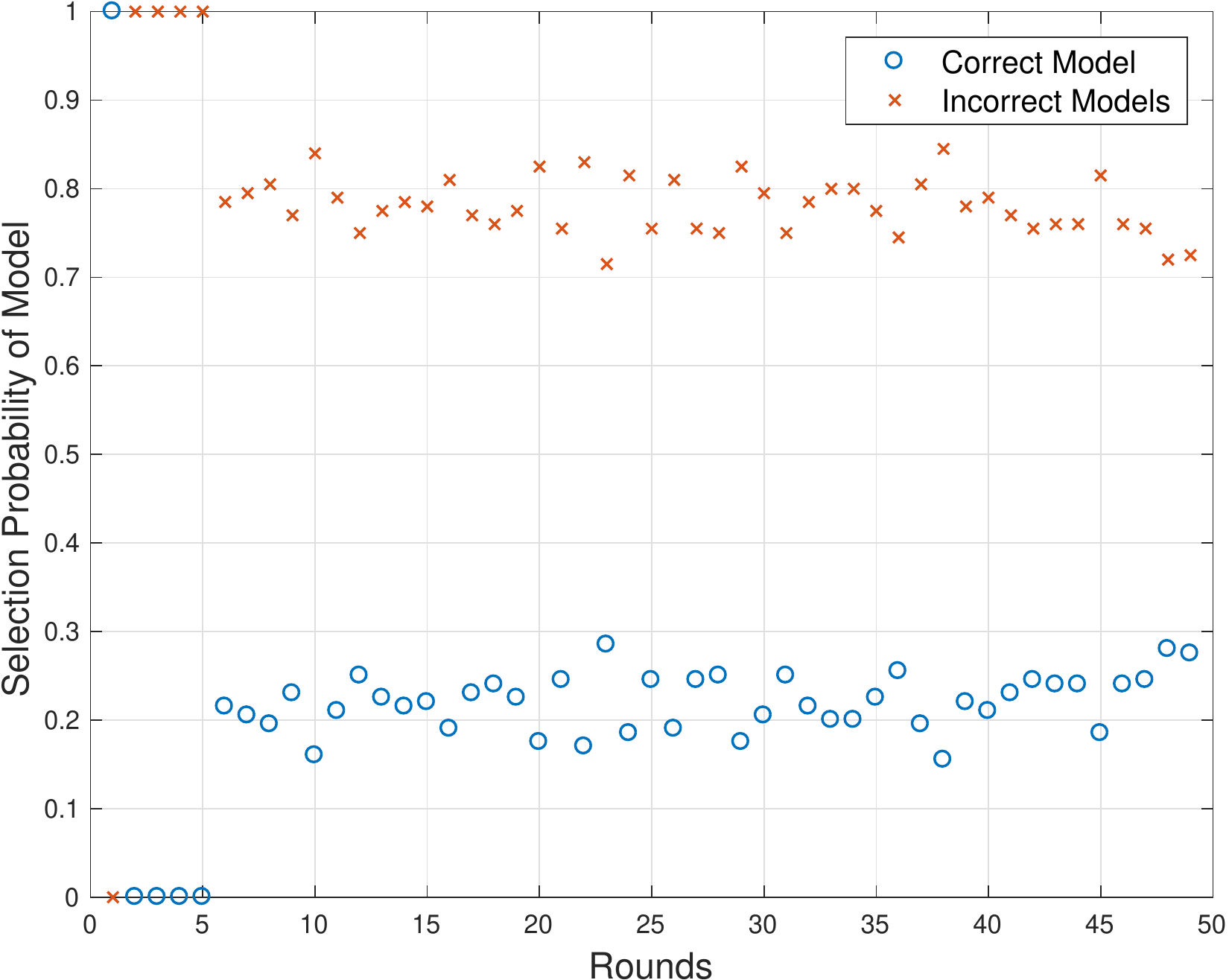} \\
(b) \\
\end{center}
\caption{Performance of model selection with Gaussian DAS
when $(K,p,N) = (100, 0.2, 4)$, $\tau = 20$
for softmax, and the correct
model is model 1: (a) MSE per round; (b) Selection probability of model
($\circ$ marks for the probability of selecting model 1
and $\times$ marks for the probability of selecting
incorrect models (i.e.,  $m \in \{2, \ldots, M\}$).}
        \label{Fig:x_plt2}
\end{figure}

Note that as mentioned earlier, the probabilities
of model selection become equal as $\tau$ increases, which 
means that the correct model cannot be selected
with a high probability.
This results in a poor performance. To see this,
we set $\tau$ to 20 and present
simulation results in Fig.~\ref{Fig:x_plt2}
with $(K,p,N) = (100, 0.2, 4)$.
It is shown that the MSE per round decreases
with rounds thanks to more measurements.
However, incorrect models are frequently chosen
due to a large $\tau$ as shown in Fig~\ref{Fig:x_plt2} (b)
(it is shown that the probability that the correct model
is chosen is about $\frac{1}{M} = 0.2$).
As a result, the performance with $\tau = 1$
(which is shown in Fig.~\ref{Fig:x_plt1} (a)) 
is better
than that with $\tau = 20$
(which is shown in Fig.~\ref{Fig:x_plt2} (a)).
Clearly, from Figs.~\ref{Fig:x_plt1} and ~\ref{Fig:x_plt2},
it is seen that the multi-armed bandit
algorithm plays a crucial role in the model selection
with DAS.
That is, a good multi-armed bandit algorithm
can help choose a right model and result in
a good estimate of $\bx$ through DAS when a correct model
is unknown.

\section{Concluding Remarks}	\label{S:Conc}

DAS was studied when
measurements at sensor nodes are assumed to be
correlated Gaussian in this paper.
It was shown that Gaussian DAS can have an 
optimal pre-determined polling order 
for nodes to sense and upload their measurements
under MMSE criterion.
However, when some nodes are unable to sense
their measurements or transmit due to deep fading,
it was shown that the polling order
has to be adaptive 
and the next node for sensing/uploading
is determined by the available data set in each round
(as expected in DAS).

Gaussian DAS was generalized when multiple parallel channels
are available. It was shown that random access
can perform better than polling when the uploading probability
is less than $e^{-1}$, which might be a typical 
case due to fading with a limited transmit power at nodes.
Another generalization was studied when multiple models
exist and the correct model is not known 
in advance. The problem was formulated as a multi-armed
bandit problem and a well-known bandit algorithm
was applied. Simulation results showed that
a bandit algorithm can be used for 
the model selection in conjunction with DAS
(so that the BS can not only have a good estimate
of measurements, but also decide the correct model
with a high probability).

\bibliographystyle{ieeetr}
\bibliography{sensor}

\end{document}